\documentclass[conference]{IEEEtran}
\IEEEoverridecommandlockouts
\usepackage{cite}
\usepackage{amsmath,graphicx}
\usepackage{url,times} 
\usepackage{array} 
\usepackage{booktabs} 
\usepackage{pifont} 
\usepackage{colortbl} 
\usepackage{tabularx}

\begin{document}
\begin{sloppy}

\title{A Hybrid Approach with Multi-channel I-Vectors and Convolutional Neural Networks for Acoustic Scene Classification}

\author{\IEEEauthorblockN{Hamid Eghbal-zadeh\IEEEauthorrefmark{1},
Bernhard Lehner\IEEEauthorrefmark{1},
Matthias Dorfer\IEEEauthorrefmark{1} and 
Gerhard Widmer\IEEEauthorrefmark{1}}
\thanks{This work was supported by the Austrian Science Fund
(FWF) under grant no. Z159 (Wittgenstein Award) and by the Austrian Ministry for Transport, Innovation and Technology,
the Ministry of Science, Research and Economy, and the Province of
Upper Austria in the frame of the COMET center SCCH.
We also gratefully acknowledge the support of NVIDIA Corporation with the donation of a Titan X GPU used for this research.
}
\IEEEauthorblockA{\IEEEauthorrefmark{1}Department of Computational Perception, Johannes Kepler University of Linz, Austria\\
Email: first.last@jku.at}}

\maketitle

\begin{abstract}
In Acoustic Scene Classification (ASC) two major approaches have been followed  .
While one utilizes engineered features such as mel-frequency-cepstral-coefficients (MFCCs), the other
uses learned features that are the outcome of an optimization algorithm.
I-vectors are the result of a modeling technique that usually takes engineered features as input. It has been shown that standard MFCCs
extracted from monaural audio signals lead to i-vectors that exhibit poor performance, especially on indoor
acoustic scenes.
At the same time, Convolutional Neural Networks (CNNs) are well known for their ability to learn features
by optimizing their filters. 
They have been applied on ASC and have shown promising results.
%The first category benefits from engineered features such as MFCCs and the second category uses feature learning techniques 
%such as Convolutional Neural Networks (CNNs).
%In contrast, the other category uses feature learning where the features are learned by an optimization algorithm.
%I-vector features are one of the feature modeling techniques that benefit from the engineered features in ASC.
%It has been shown that 
%when using standard MFCCs extracted from mono audios, 
%i-vector based approaches suffer from the poor performances on indoor-scenes.
%
%Convolutional Neural Networks (CNNs) are a well known feature learning method that optimize their filters to learn their own feature.
%They have been applied on ASC and have shown promising results.
%
In this paper, we first propose a novel multi-channel i-vector extraction scheme for ASC, improving their performance on indoor and outdoor scenes.
Second, we propose a CNN architecture that achieves promising ASC results.
Further, we show that i-vectors and CNNs capture complementary information from acoustic scenes.
%The purpose of this paper is threefold. 
%First to improve the i-vector representation for ASC.
%We propose a novel multi-channel i-vector extraction scheme to improve the indoor-scene as well as outdoor-scene predictions
%for i-vector based approaches in ASC.
%
%Second, we show that the features learned via CNNs contain complementary information to the i-vector features.
%we show that although both CNNs and i-vectors can achieve promising results, the information captured with each method 
%differs from the other.
%
%Finally, we provide an efficient way to combine the i-vector features and CNNs for ASC.
Finally, we propose a hybrid system for ASC using multi-channel i-vectors and CNNs by utilizing a score fusion technique.
%
%For our experiments, the TUT16 dataset is used and our system is compared with the state-of-the-art in ASC.
%The experimental results suggest that our hybrid approach achieves better performances compared to the other ASC methods.
Using our method, we participated in the ASC task of the DCASE-2016 challenge.
Our hybrid approach achieved $1^{\text{st}}$ rank among 49 submissions, substantially improving the previous state of the art.
\end{abstract}

% \fi
\IEEEpeerreviewmaketitle
\section{Introduction}
The field of Computational Auditory Scene Analysis (CASA) is a fast moving area, with international challenges such as DCASE~\footnote{\url{www.cs.tut.fi/sgn/arg/dcase2016/}} to accelerate the progress on complex audio recognition problems. 
%
%Recent advances combining different data modelling and machine learning approaches -- e.g.deep learning~\cite{piczak2015environmental}, matrix factorization ~\cite{bisot2016acoustic} and factor analysis~\cite{elizalde2013vector} are pushing the state of the art in audio scene classification.
%
Previously, methods based on feature engineering from speech and music domain such as 
MFCCs, Linear Predictive Coefficients and Gammatone Cepstral Coefficients\cite{valero2012gammatone} have been used to extract features for ASC.

More sophisticated techniques have been developed to overcome the complexities caused by the noise and other unwanted acoustic phenomena.
Such methods manipulate the engineered feature spaces to create a better acoustic representation to distinguish between different acoustic scenes.
For example, i-vector features\cite{dehak2011front} use MFCCs to create a low-dimensional latent space for short audio segments. 

In contrast with feature engineering techniques, there exist methods that learn an internal representation from spectograms or similar representations 
of audio by optimizing their parameters.
Examples of such methods are Non-negative Matrix Factorization (NMF)~\cite{bisot2016acoustic} and CNNs~\cite{salamon2016deep}.
In the scientific society there are many discussions about the use of feature engineering approaches versus feature learning methods.
The authors of this paper believe that CASA could benefit from the best of both worlds. 
Therefore, we introduce a hybrid approach for ASC.
The aim of this work is threefold:
\begin{enumerate}
\item We introduce a novel multi-channel i-vector extraction scheme for ASC using tuned MFCC features 
extracted from both channels of audio.
%We provide experimental results, showing our novel scheme for i-vector extraction significantly improves both indoor  
%and outdoor scene predictions which leads to an overall performance improvement for ASC.
\item Using a CNN architecture inspired by \textit{VGG style} networks 
%borrowed from image object recognition, 
we show that we can achieve comparable performances to the state of the art in ASC.
Also, we show that i-vectors and CNNs provide complementary information from the acoustic scenes.
%Also we show that the information CNNs capture by learning their internal representation, differs from what i-vector features capture.
%Therefore, a hybrid method consist of the two would be beneficial for an ASC system.
%We provide the experimental results showing our CNN can achieve promising results in classifying audio scenes.
\item Finally, we demonstrate an efficient way of fusing i-vectors with CNNs.
%By comparing class-wise performances we show that i-vectors and CNNs have different weaknesses and strengths, 
We propose a hybrid ASC system using multi-channel i-vectors, CNNs, and  a score-fusion technique.
\end{enumerate}
%In this paper, 
%
%
%We further examine a score calibration technique with our multi-channel i-vector system.
%Third, we use a Deep Convolutional Neural Network (DCN) trained
%on spectrograms of audio excerpts in an end-to-end fashion.
%Finally, we propose a hybrid system which benefits from a late-fusion of the binaural i-vector and the DCN systems.
%Using our proposed methods, we ranked \textbf{first} and \textbf{second} among 49 submissions in the audio scene classification task of DCASE-2016 challenge.
%
%The reminder of this paper is organized as follows. 
%In the Section~\ref{sec:ivec}, the i-vector representation is described.
%Our novel multi-channel i-vector extraction scheme is explained in Section~\ref{sec:BMB}.
%The DCN approach is detailed in Section~\ref{sec:DCN}.
%The score post-processing techniques are described in Section~\ref{sec:score_proc}.
%In Section~\ref{sec:eval}, the results are presented.
%Finally, Section~\ref{sec:conclude} concludes this paper.
%
%
\section{I-vector Features}
\label{sec:ivec}
%\subsection{Theory}
The i-vector~\cite{dehak2011front} representation has been used in different areas such as speech~\cite{zeinali2016deep,bahari2013accent}, music classification~\cite{eghbalzISMIR2015ivecsim} and ASC~\cite{elizalde2013vector}.
I-vectors are segment-level representations computed from acoustic features (e.g., MFCCs) of audio segments.
They provide a fixed-length and low-dimensional representation for audio excerpts containing rich acoustic information.
In short, i-vectors are latent variables representing the shift of an audio segment from a universal distribution, usually referred to as Universal Background Model (UBM).

The UBM is trained on the acoustic features of a sufficient amount of audio segment examples to capture the distribution of the acoustic feature space.
To extract the i-vector of an audio segment, the UBM model is first adapted to acoustic features of the segment and the parameters of the adapted model are considered as a new representation for the audio segment.

Second, to capture the shift of the adapted model from the UBM a Factor Analysis (FA) procedure is applied.
The parameters of the adapted model are decomposed into a factor with lower dimensioality and better discrimination power.
The i-vector itself, is a maximum a postriori (MAP) estimation of this low-dimensional factor.

The UBM is usually a Gaussian Mixture Model (GMM) trained on MFCCs.
To apply the FA, the adapted GMM mean supervector $\mathbf{M_s}$ -- which is adapted to an audio segment from the scene $s$ -- can be decomposed as follows:

\begin{equation}
\mathbf{M_s} = m + \mathbf{T} . y_s
\end{equation}

where $m$ is the UBM mean supervector and $\mathbf{T} . y_s$ is an offset that captures the shift from the UBM. 
The low-dimensional vector $y_s$ is a latent variable with a normal prior and its respective i-vector is a MAP estimate of $y_s$. 
The factorization matrix $\mathbf{T}$ is learned via expectation maximization.
%More information about the training procedure of $\mathbf{T}$ and i-vector extraction can be found in \cite{dehak2011front,kenny2005joint}.

\subsection{The I-vector pipeline}
To use the i-vector features for ASC, we apply a series of processing blocks known as \emph{i-vector pipeline}.
Our i-vector pipeline consists of 3 phases: 1) development, 2) training and 3) testing.
During the \emph{development phase}, the UBM is trained and the adapted models for each audio segment in the development set are computed.
Then using these adapted models, the $\mathbf{T}$ matrix is trained.

In the \emph{training phase}, i-vectors of the training set are extracted by utilizing the i-vector models (UBM and $\mathbf{T}$) and the 
length of i-vectors are normalized to one~\cite{garcia2011analysis}.
Using these training set i-vectors, a Linear Discriminant Analysis (LDA)~\cite{scholkopft1999fisher} and a Within-Class Covariance Normalization (WCCN)~\cite{hatch2006within} model is trained. %WCCN reduces the within-class covariance by projecting the i-vectors in the opposite direction of within-class covariance matrix.
The LDA and WCCN projection matrices are used for projecting i-vectors in order to reduce the within-class variability 
and maximize the class separation in the i-vector space.
Afterwards, the class-averaged i-vectors are stored as the \emph{model i-vector} for each class.

In the \emph{testing phase}, the i-vectors are extracted, length-normalized and projected by using the previously trained models
(UBM, $\mathbf{T}$, LDA, WCCN). 
Each resulting i-vector from the test set is then compared to all the model i-vectors by applying cosine scoring~\cite{dehak2010cosine}. 
Finally, the class represented by its model i-vector with the highest score, determines the predicted class. 
 
%\subsection{Our I-vector setup} 
Our UBM is trained with 256 Gaussian components on MFCC features extracted from audio excerpts. 
The UBM, $\mathbf{T}$ matrix, LDA and WCCN projections are trained on the training portion of
each Cross Validation (CV) split. 
The dimensionality of the i-vector space is set to 400. 
The configuration of MFCC features are discussed in the following section.
%The details of our MFCC features are explained in the following section. 
%\subsection{Our I-vector setup}
%We trained our UBMs with 256 Gaussian components on MFCC features extracted from audio excerpts.
%The UBM, $\mathbf{T}$ matrix, LDA and WCCN projections are trained on the training portion of each cross-validation split.
%The details of our MFCC features are explained in the following section.
%We set the dimensionality of the i-vectors to 400.

%For I-vector post-processing, we projected i-vectors via Linear Discriminant Analysis,
%followed by Within-Class Covariance Normalization (WCCN).
%To score the projected i-vectors, we used a cosine scoring as explained in~\cite{dehak2010cosine}.

%\begin{figure}[t]
%  \centering
%  \centerline{\includegraphics[width=\columnwidth]{figs/ivec_pipeline.eps}}
%  \caption{Block-diagram of the i-vector pipeline.}
%  \label{fig:diag}
%\end{figure}

%
% Section by Bernhard
%
\section{Multi-channel I-Vector Extraction Scheme}
\label{sec:BMB}
%In this section, we present our multi-channel i-vector extraction scheme. 
To improve the performance of i-vectors for ASC, we propose a novel multi-channel i-vector extraction scheme.
Our scheme can be explained in 3 steps:
1) MFCC parameter tuning, 2) multi-channel i-vector extraction and 3) score averaging.
We first demonstrate a parametrization of MFCC features, tuned for i-vector extraction for ASC.
Further, we explain how we use the multi-channel signals for i-vector extraction.
Finally, we describe the score averaging technique.% used for ASC. 

\subsection{Boosting MFCCs for i-vector extraction}
\label{sec:intro}

In~\cite{lehner2011mfcc} it was shown that it is useful to find a good
parametrisation of MFCCs for a given task. Therefore, the first step is to
improve the performance of MFCCs which we extract with the Voicebox toolbox\cite{Voice99}.
%
%In order to include all the components that are involved, we do this after the complete i-vector pipeline is implemented. 
The results in this section are always averaged from a four-fold CV.%, unless explicitly mentioned otherwise.

%\subsubsection{Investigating the Windowing Scheme}
%We investigate different observation window
%lengths to find the optimum point in MFCC feature extraction for i-vector computation.
%For this, 
To investigate different observation window lengths, we place the different observation windows symmetrically around the frame that was always
fixed on 20 ms. 
Thus, independent of the actual observation window, we always
end up with exactly the same amount of observations.
In Table \ref{tab:mfcc_tune} we provide the results of different windowing
schemes for MFCCs and their deltas and double deltas. 
As can be seen, the impact of using different overlaps is quite severe on the
results of the MFCCs. Experimental results suggest that a 20 ms window without overlap gives
best accuracy using i-vectors extracted from MFCCs.
The effect is much smaller on the results of deltas and double deltas.
Nevertheless, we consider it useful to
extract deltas and double deltas separately with a 60 ms observation window,
and combine them with the 20 ms MFCCs into one single feature vector.

\begin{table}
 \begin{center}\setlength{\tabcolsep}{5pt}
 \begin{tabular}{l|lll|ll}
  \textbf{(\%) } &  \multicolumn{3}{c|}{window length}& \multicolumn{2}{c}{coefficients}
  \\\hline
  \textbf{ } & win=20 ms  & win=60 ms  & win=100 ms&w/ $0^{th}$ &w/o $0^{th}$
  \\
  \hline
  $MFCC$ & \cellcolor[RGB]{220,220,220}68.95 & 61.84 & 60.61&68.95 & \cellcolor[RGB]{220,220,220}71.43
  \\
  $\Delta$ & 61.62  & \cellcolor[RGB]{220,220,220}64.02 & 60.68 &\cellcolor[RGB]{220,220,220}61.62 & 56.34
  \\
  $\Delta\Delta$ & 61.54 & \cellcolor[RGB]{220,220,220}62.05 &59.49 &\cellcolor[RGB]{220,220,220}61.54 & 50.77
  \\
  \hline
 \end{tabular}
\end{center}
 \caption{Results of MFCC tuning. Gray cells indicate the configurations that were combined
 for further experiments.}
 \label{tab:mfcc_tune}
\end{table}

%Regardless of the length of the observation window, MFCCs
% without the 0th coefficient consistently perform better.

%\subsubsection{Investigating the Number of Coefficients}
After fixing observation window lengths for MFCCs and deltas and double deltas,
we evaluate the amount of coefficients that is actually useful in our specific
setting. Often, the $0^{th}$ coefficient is ignored in order to achieve
loudness invariance, which is also helpful for ASC.
The results in Table \ref{tab:mfcc_tune} support this intuition, where we can
see that including the $0^{th}$ coefficient leads to reduced accuracy for i-vectors extracted from MFCCs. 
%($\approx 71\%$ vs. $\approx 69\%$).
%
Nevertheless, including the delta and double delta of the $0^{th}$
MFCC is in the feature vectors shows performance improvement as shown in Table \ref{tab:mfcc_tune}.
% it can be seen that the performance of the i-vectors extracted from MFCC deltas drops from $61.6\%$ to $56.3\%$
%accuracy without the $0^{th}$ MFCC deltas. The performance of the i-vectors extracted from MFCC double
%deltas drops from $61.5\%$ to $50.8\%$ without the $0^{th}$ MFCC double deltas.
%
%\begin{table}
% \begin{center}\setlength{\tabcolsep}{6pt}
% \begin{tabular}{ll|ll|ll}
%  %\hline
%  \multicolumn{2}{c|}{$MFCC$}
%  & \multicolumn{2}{c|}{$\Delta$}
%  & \multicolumn{2}{c}{$\Delta\Delta$}
%  \\
%  w/ $0^{th}$ & w/o $0^{th}$ & w/ $0^{th}$ & w/o $0^{th}$ & w/
%  $0^{th}$ & w/o $0^{th}$
%  \\
%  \hline
%  68.95 & 71.43  & 61.62 & 56.34 & 61.54 & 50.77
%  \\
%  %\hline
%  \hline
% \end{tabular}
%\end{center}
% \caption{The impact of the $0^{th}$ MFCC. It can be seen that it is important
% to include just the $0^{th}$ MFCC deltas and double deltas, but not the
% $0^{th}$ MFCC itself.}
% \label{tab:0th_coeff}
%\end{table}
%
%
%In a series of further experiments conducted, the amount of coefficients that
%turned out to be useful was determined, separately for MFCCs, deltas and double
%deltas.
%
%\subsubsection{Final MFCC configuration}

The configuration used to extract \emph{MFCC boosted i-vectors} are as follows.
We use 23 MFCCs (without $0^{th}$ MFCC) extracted by applying a 20 ms
observation window without any overlap. 18 MFCC deltas (including the $0^{th}$ MFCC
delta), and 20 MFCC double deltas (including the $0^{th}$ MFCC double delta) are
extracted by applying a 60 ms observation window, placed symmetrically around a
20 ms frame.
Regardless of the observation window length, we use 30 triangle shaped
mel-scaled filters in the range [0-11 kHz].

\subsection{Multi-channel Feature Extraction}
Most often, the binaural audio material is down-mixed into a single monaural
representation by simply averaging both channels. 
This could be problematic in cases where an important cue is only captured well in one of the channels, since averaging would then lower the SNR, and increase the chance that it gets missed by the system. The analysis of both channels separately would alleviate this problem.

Not only do we extract MFCCs from both channels separately, but also from the
averaged monaural representation as well as from the difference of both
channels. 
All in all, we extract MFCCs from four different audio sources,
resulting in four different feature space representations per audio file. 
An experiment where we concatenated the MFCCs into a single feature vector did not
lead to improved i-vector representations, therefore we opt for a score-averaging approach.

\begin{table}
 \begin{center}\setlength{\tabcolsep}{6.5pt}
 \begin{tabular}{l|llllll}
  %\hline
%%  \textbf{ } & acc [\%] & f1 [\%] & acc [\%] & f1 [\%] & acc [\%] & f1 [\%]
  \textbf{(\%) } & fold 1 & fold 2 & fold 3 & fold 4 & avg
  \\
%%  \textbf{ } & acc & acc & acc & acc & acc
  %%\\
  \hline
%  %74.1 %   |  |  79.7 %  |  64.3 %  |  77.1 %  |  75.3 %  
%BASE & 79.7 & 64.3 & 77.1 & 75.3 & 74.1 
BAS&78.97 &64.48 & 68.46& 77.5&72.24
  \\
  %Monaural& 85.52 & 65.86 & 79.19 & 77.05 & 76.91 
	Single-ch.& 84.14&68.28&75.17&75.00& 75.65
  \\
  Multi-ch.& \cellcolor[RGB]{220,220,220}85.86 & \cellcolor[RGB]{220,220,220}76.55 & \cellcolor[RGB]{220,220,220}77.52 & \cellcolor[RGB]{220,220,220}83.22 & \cellcolor[RGB]{220,220,220}80.79
  \\
%%  $MFCC III$ & 87.93 & 80.34 & 81.54 & 85.62 & 83.86
%%  \\
%%  $CNN$ & 80.69 & 75.52 & 77.85 & 83.90 & 79.49
%%  \\ 
%%  $COMB$ & 94.48 & 85.17 & 87.25 & 92.81 & 89.93
%%  \\
  \hline
  %\hline
 \end{tabular}
\end{center}
\caption{Comparing the performance of the i-vector system that uses our tuned MFCCs, with the provided MFCCs.
 BASE: original MFCC provided by the DCASE organisers; Single-channel: tuned MFCCs on averaged single-channel;
 multi-channel: multi-channel tuned MFCCs.}
 \label{tab:final_opt}
\end{table}

%\subsection{Score Post-Processing}
%\label{sec:score_proc}
%In order to further improve the performance of our system, we suggest 
%the following score post-processing techniques.

%\subsection{Score Averaging}
The aforementioned separately extracted MFCCs yield four different i-vectors and LDA-WCCN models
which in turn result in four different cosine scores per audio file. In order to fuse those scores, we simply compute the mean of them and use it as the final score for each audio excerpt.
%Please note that the score averaging after the scoring step totally has a different effect than averaging the signal into a monaural audio for the modeling.
%In there, the different acoustic events from different channels are already captured separately and are projected into the i-vector space.
%The score-averaging allows the i-vector of each channel to express the acoustic event that it is captured by increasing its score.
%This can be presented later on, in the average of the scores.
%While by averaging the signal into monaural, any opportunity of capturing the acoustic events in different channels are already lost in the signal level.
%Additionally, we utilise a score averaging approach for the unseen Test
%set, where we combine the output of the models trained on the four CV folds by averaging.
%All in all, we combine 16 scores in order to yield the classification result of
%one audio file.
%\subsection{Calibration} 
%
% 
% Voicebox 
% 
% Optimised MFCCs
% \cite{lehner2011mfcc}
% 
% MSR Identity toolbox \cite{MSRIDToolbox2013}
%
\section{Deep Convolutional Neural Networks}
\label{sec:dcnn}
As mentioned in the introduction, one of the contributions of this work is to study the differences between approaches using feature engineering and feature learning in ASC.
%we describe our CNN architecture used for feature learning from the spectrograms of acoustic scene recordings.
%We designed a CNN architecture for ASC. 
In this section, we describe the neural network architecture as well as the optimization
strategies used for feature learning in our ASC network.

The specific network architecture used is depicted in Table \ref{tab:model_architecture}.
The feature learning part of our model follows the \textit{VGG style networks} for object recognition and the classification part
of the network is designed as a global average pooling layer
as in the \textit{Network in Network} architecture\cite{lin2013network}.
The input size of our network is a one channel spectrogram excerpt with size $149 \times 149$.
This means we train the model not on whole sequences but only on small "sliding" windows.
The spectrograms for this approach are computed as follows:
The audio is sampled at a rate of $22050$ samples per second.
We compute the Short Time Fourier Transform (STFT) on $2048$
sample windows at a frame rate of $31.25$ FPS.
Finally we post-process the STFT with a logarithmic filterbank with 24 bands, logarithmic magnitudes
and an allowed passband of $20$Hz to $16$kHz.
The parameters of our models are optimized with mini-batch stochastic gradient decent and momentum.
The mini-batch size is set to $100$ samples.
We start training with an initial learning rate of $0.02$ and half it every 5 epochs.
The momentum is fixed at $0.9$ throughout the entire training.
In addition we apply an $L2$-weight decay penalty of $0.0001$ on all trainable parameters of our model.

For classification of unseen samples at test time we proceed as follows.
First we run a sliding window over the entire test sequences and collect
the individual class probabilities for each of the window.
In a second step we average the probabilities of all contributions and assign the class
with maximum average probability.

\begin{table}[t]
\label{tab:model_architecture}
\begin{center}
\begin{tabular}{c}
Input $1 \times 149 \times 149$ \\
\hline
$5\times5$ Conv(pad-2, stride-2)-$32$-BN-ReLu \\
$3\times3$ Conv(pad-1, stride-1)-$32$-BN-ReLu \\
$2\times2$ Max-Pooling + Drop-Out($0.3$) \\
\hline
$3\times3$ Conv(pad-1, stride-1)-$64$-BN-ReLu \\
$3\times3$ Conv(pad-1, stride-1)-$64$-BN-ReLu \\
$2\times2$ Max-Pooling + Drop-Out($0.3$) \\
\hline
$3\times3$ Conv(pad-1, stride-1)-$128$-BN-ReLu \\
$3\times3$ Conv(pad-1, stride-1)-$128$-BN-ReLu \\
$3\times3$ Conv(pad-1, stride-1)-$128$-BN-ReLu \\
$3\times3$ Conv(pad-1, stride-1)-$128$-BN-ReLu \\
$2\times2$ Max-Pooling + Drop-Out($0.3$) \\
\hline
$3\times3$ Conv(pad-0, stride-1)-$512$-BN-ReLu \\
Drop-Out($0.5$) \\
$1\times1$ Conv(pad-0, stride-1)-$512$-BN-ReLu \\
Drop-Out($0.5$) \\
\hline
$1\times1$ Conv(pad-0, stride-1)-$15$-BN-ReLu \\
Global-Average-Pooling \\
\hline
$15$-way Soft-Max\\ \\
\end{tabular}
\caption{Model Specifications. BN: Batch Normalization, ReLu: Rectified Linear Activation Function, CCE: Categorical Cross Entropy. For training a constant batch size of 100 samples is used.}
\end{center}
\end{table}

%
% Section by Hamid
%
\section{Score Fusion}
\label{sec:fuse}
To fuse the multi-channel i-vector cosine scores with the soft-max activation probabilities of our VGG-net, 
a score fusion technique is carried out.
We use a Linear Logistic Regression (LLR) model to learn the fusion parameters.
Our LLR learns the coefficients $\alpha_k$ and bias $\beta$ to fuse the likelihoods of $K$ different models by maximizing $C^\prime_{llr}$ where:
\begin{equation}
\begin{split}
 {\vec{l^\prime}(x_n)}=\sum\limits_{k =1}^K\alpha_k \cdot \vec{l_k}(x_n) + \beta\\
C^\prime_{llr} = - \frac{1}{N}\sum\limits_{n =1}^N w_c \log P^\prime_n
\end{split}
\end{equation}
and $C$ is the number of our classes and $N$ is the total number of our likelihood examples.
$\vec{l_k}(x_n)$ is the $n^{\text{th}}$ likelihood from the model $k$, $P^\prime_n$ is the posterior probability for the true class of $n^{\text{th}}$ example, 
given the fused likelihood $\vec{l^\prime}(x_n)$ and a flat prior.
The flat prior is $\frac{1}{C}$ where $c$ is the true class of $n^{\text{th}}$ example.
$w_c$ is the ratio between number of classes, 
and the number of examples available from each class defined as $\frac{\frac{1}{C}}{\frac{N_c}{N}}$ where $N_c$ is the number of available likelihoods from the class $c$.
The parameters of LLR are learned using the scores on the validation set and applied on the test set scores of different models for the final fusion.
Our score fusion has two purposes: 1) to fuse the i-vector cosine scores with CNN probabilities, and 2) to calibrate the scores and 
reduce the score distribution mismatch between the training and validation scores, as explained in~\cite{brummer2007focal}.
\footnote{In our CMB system, only the averaged cosine scores of validation set are used to train the fusion model.
Therefore the fusion model for CMB, only works as a score calibration model.}
%The LLR is trained in a way to map the cosine scores and CNN scores of the training set into a one The fused scores are used for the final prediction.
For the score fusion, the FoCal Multi-class toolkit is used~\cite{brummer2007focal}.
To fuse the scores on evaluation set, we follow a bootstrap aggregating\cite{breiman1996bagging} approach and combine the output of fusion models already trained on the four CV folds by averaging the fused scores on each fold.
\section{Evaluation}
\label{sec:eval}
To evaluate the performance of different methods, we use TUT database for ASC (TUT16)~\cite{mesaros2016tut}.
%We use both development and evaluation sets for performance comparison using the accuracy.
On the development set, we follow a four-fold CV provided with the dataset.
On the evaluation set, we train on the development set and test on the evaluation set.
The performance of all the methods on the evaluation set are taken from the DCASE-2016 challenge results and their respective articles available on the challenge website.
\subsection{Baselines}
\label{subsec:baselines}
%We use four baselines to compare our proposed methods with the other approaches in ASC.
Our first baseline is a GMM, trained on monaural standard MFCCs\cite{mesaros2016tut}, 
which is similar to the i-vector paradigm but lacks the factor analysis step.
%This baseline can be found as \emph{GMM} in the results section.
Our second baseline is a supervised NMF\cite{Bisot2016}.
%The results of this method are shown under \emph{NMF}.
Our third baseline is a CNN optimized with categorical cross-entropy loss function, trained on the logarithmic conversion of the mel power of the input audios~\cite{Valenti2016}. 
%We address our second baseline as \emph{CNN} in the results.
Our fourth baseline (BAS) is an i-vector system based on our pipeline, using monaural standard MFCCs extracted with TUT16 baseline implementation.
%
%The proposed methods also can be found as SMB (Single-channel MFCC Boosted I-vectors, Section~\ref{sec:BMB}), MMB (Multi-channel MFCC Boosted I-vectors, Section~\ref{sec:BMB}) and CMB (Calibrated Multi-channel MFCC Boosted I-vectors, Section~\ref{sec:BMB}).
In order to demonstrate the impact of each step in our multi-channel i-vector extraction scheme, we report the 
results on single-channel MFCC boosted i-vectors (SMB), on multi-channel MFCC boosted i-vectors (MMB) and on calibrated multi-channel MFCC boosted i-vectors (CMB).
The results of our VGG-net and our final hybrid system can be found as (VGG) and (HYB) in the results section, respectively.
%We provided 4 different submissions based on the methods described in the previous sections:
%\begin{enumerate}
 %  \item DCN: Deep Convolutional Neural Network (explained in Section~\ref{sec:DCN})
%	\item BMB: Binaural MFCC Boosted I-vectors (explained in Section~\ref{sec:BMB})
%	\item CBMB: Calibrated Binaural MFCC Boosted I-vectors (explained in Section~\ref{sec:BMB})
%	\item LFBI: Late Fusion of CNN and Binaural I-vectors (explained in Section~\ref{sec:fuse})
%\end{enumerate}
%
%\subsection{Results}
%\label{subsec:results}
%The discussions on the experimental results are available in the following section.
%
%
\begin{table}[]
\centering
\begin{tabular}{l|llllll}
 (\%)        &BAS& SMB & MMB     & CMB   &VGG   &HYB \\ \hline
\cellcolor[RGB]{173,216,230}Beach        &83.52&76.82& 78.95 & 86.84 &\cellcolor[RGB]{173,216,230}92.11 &\cellcolor[RGB]{144,238,144}92.11  \\
\cellcolor[RGB]{255,160,122}Bus          &68.16&74.67& 79.47 &\cellcolor[RGB]{255,160,122} 87.11 &77.37 &\cellcolor[RGB]{144,238,144}95.00 \\
\cellcolor[RGB]{173,216,230}Cafe/Rest.   &66.68&55.24& 62.87 & 78.72 &\cellcolor[RGB]{173,216,230}80.27 &\cellcolor[RGB]{144,238,144}93.92 \\
\cellcolor[RGB]{255,160,122}Car          &64.93&96.18 & 96.18 & \cellcolor[RGB]{255,160,122}96.18&84.61 &\cellcolor[RGB]{144,238,144}96.18\\
\cellcolor[RGB]{255,160,122}City         &84.10&86.78 & 90.19 & \cellcolor[RGB]{255,160,122}90.01&83.79 &88.52\\
\cellcolor[RGB]{255,160,122}Forest       &82.94&93.65& 94.84 & \cellcolor[RGB]{255,160,122}96.03 &94.05 &\cellcolor[RGB]{144,238,144}98.81\\
\cellcolor[RGB]{173,216,230}Groc. Store  &70.61&89.60& 94.86 & 89.72 &\cellcolor[RGB]{173,216,230}93.80 &\cellcolor[RGB]{144,238,144}95.11  \\
\cellcolor[RGB]{173,216,230}Home         &82.87&55.38& 59.15 & 71.01 &\cellcolor[RGB]{173,216,230}72.29 &\cellcolor[RGB]{144,238,144}89.17\\
\cellcolor[RGB]{255,160,122}Library      &61.76&72.52& 75.56 & \cellcolor[RGB]{255,160,122}78.13 &75.14 &\cellcolor[RGB]{144,238,144}85.93 \\
\cellcolor[RGB]{173,216,230}Metro        &95.98&77.37& 83.92 & 84.10 &\cellcolor[RGB]{173,216,230}88.52 &\cellcolor[RGB]{144,238,144}91.89 \\
\cellcolor[RGB]{255,160,122}Office       &78.23&93.06& 97.22 & \cellcolor[RGB]{255,160,122}90.50 &73.18 &\cellcolor[RGB]{144,238,144}97.22 \\
\cellcolor[RGB]{255,160,122}Park         &35.42&64.03& 78.33 & \cellcolor[RGB]{255,160,122}81.81 &58.61 &\cellcolor[RGB]{144,238,144}86.94 \\
\cellcolor[RGB]{255,160,122}Resid. Area  &70.55&45.68& 63.60 & \cellcolor[RGB]{255,160,122}72.06 &67.54 &\cellcolor[RGB]{144,238,144}76.00  \\
\cellcolor[RGB]{255,160,122}Train        &48.85&71.63 & 73.18 & \cellcolor[RGB]{255,160,122}72.95&63.45 &\cellcolor[RGB]{144,238,144}76.74 \\
\cellcolor[RGB]{173,216,230}Tram         &89.52&85.74& 86.99 & 84.47 &\cellcolor[RGB]{173,216,230}90.66 &87.88\\ %\\%\hline
\end{tabular}
\caption{Class-wise accuracies of different methods for ASC on the development set of TUT16 dataset. The results are averaged over all the folds. Each class is painted with a color representing the better performing model (CMB:red, VGG:blue). Green shows where HYB performed best.}
\label{tab:class_wise}
\end{table}
\begin{figure}[t]
  \centering
  \centerline{\includegraphics[width=\columnwidth]{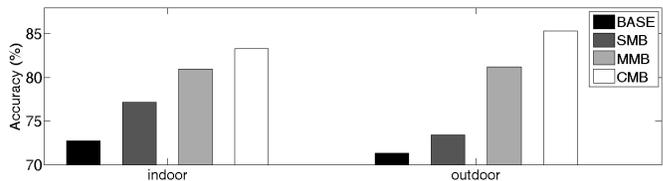}}
  \caption{Performance comparison of different methods for indoor and outdoor scenes on development dataset of TUT16.}
  \label{fig:in_out}
\end{figure}
\begin{table}[]
\centering
\begin{tabular}{p{0.4cm}|p{0.4cm}p{0.4cm}p{0.4cm}p{0.5cm}|p{0.4cm}p{0.4cm}p{0.4cm}p{0.4cm}p{0.4cm}}%(\%)   &
%\begin{tabular}{l|lll|llll}%(\%)   &
%\begin{tabularx}{.8\textwidth}{l|lll|llll}%(\%)   &
   &GMM&NMF&CNN&BAS&SMB&MMB&CMB&VGG&HYB\\ \hline
 eval.&77.2&87.7&86.2&-&-&86.4&88.7&83.3&\textbf{89.7}\\ \hline
 dev.&72.5 &86.2&79.0&72.24&75.65&80.8&83.9&79.5&\textbf{89.9} \\ \hline
rank&28&3&6&-&-&5&2&14&\textbf{1}\\ %\\
%\end{tabularx}
\end{tabular}
\caption{The accuracies on TUT16 dataset are provided for both evaluation set and development set. Also, we provide the ranks of each method for the DCASE-2016 challenge taken from the challenge's website.}
\label{table:test_results1}
\end{table}
\begin{table}[]
\centering
\begin{tabular}{llllll}
   (\%)   & fold1  & fold2  & fold3  & fold4  & avg    \\ \hline 
%DCN  & 80.69 & 75.52 & 77.85 & 83.90 & 79.49 \\ \hline
BAS  &78.97 &64.48 & 68.46& 77.5&72.24 \\ \hline
SMB  &84.14&68.28&75.17&75.00& 75.65\\ \hline
MMB  & 85.86 & 76.55 & 77.52 & 83.22 & 80.79 \\ \hline
CMB & 87.93 & 80.34 & 81.54 & 85.62 & 83.86 \\ \hline
VGG &   80.69 & 75.52 & 77.85 & 83.90 & 79.49  \\ \hline
HYB &  \textbf{94.48} & \textbf{85.17} & \textbf{87.25} & \textbf{92.81} & \textbf{89.93}  \\ %\\%\hline
\end{tabular}
\caption{Accuracies of different methods on the development set cross-validation splits, provided with TUT16 dataset.}
\label{tab:validation_results}
\end{table}
\section{Results and Discussion}
\label{sec:disc}
In Table~\ref{table:test_results1} we present the results of our proposed methods and the baselines on the evaluation set and development set of TUT16.
In addition, to provide a deeper insight, the performances of all the folds are reported separately in Table~\ref{tab:validation_results}.
To highlight the effectiveness of our multi-channel i-vector extraction scheme, a comparison between indoor and outdoor scene predictions are shown in Figure~\ref{fig:in_out}.
To investigate the differences between the i-vector system and our CNN, the class-wise performances are provided in Table~\ref{tab:class_wise}.
\subsection{Performance comparison}
\label{subsec:perf_comp}
As can be seen in Table~\ref{table:test_results1}, on both the development and evaluation set, HYB has a better performance compared to all the baselines and proposed methods.
Also CMB achieved the second best performance on the evaluation set and the third best performance on the development set.
Comparing BAS with SMB, MMB and CMB shows that MFCC tuning, multi-channel i-vector extraction and score calibration improves the performance of the i-vector baseline step by step. 
Specially comparing BAS and CMB demonstrates the effectiveness of our multi-channel i-vector extraction scheme, improving BAS by 16 percentage points on the development set.
This performance improvement is also visible in all the folds, as reported in Table~\ref{tab:validation_results}.
Comparing the CNN-based methods, CNN baseline performs better than our VGG-net.
One reason could be in the use of mel-energy features rather than spectrograms which gives the CNN a benefit by having a more compact representation (60 dimension) compared to our spectrograms (149 dimensions).
Between the feature learning methods (NMF, CNN and VGG), NMF achieved better performances on both development and evaluation set.
And CMB outperformed all the other feature modeling techniques that used engineered features (GMM, BAS, SMB and MMB).
%Class-wise performances shown in Table~\ref{tab:class_wise} suggests that in two cases of \textit{Home} and \textit{Tram}, BASE has a better (11.9 and 5 percentage points, respectively) class-wise performance compared to CMB.
%For all the other classes, the improvements are very noticeable and some classes such as \textit{Park}, \textit{Car} and \textit{Train} improved by 46.39, 31.3 and 24.1 percentage points respectively.

\subsection{Improving I-vector Representation}
\label{subsec:improv_iv}

As discussed in the introduction, one downside of i-vector features is their degradation in indoor scenes.
In Figure~\ref{fig:in_out} we demonstrated the overall performances for indoor scenes (\textit{Bus, Cafe, Car, Grocery store, Home, Library, Metro, Office, Train} and \textit{Tram}) and outdoor scenes (\textit{Beach, City center, Forest, Park} and \textit{Residential area}).
Comparing BAS with SMB shows that our parameter tuning step improved both indoor and outdoor scene predictions.
By looking at SMB and MMB we can observe that using multi-channel i-vector extraction scheme further improves the prediction performances.
Finally, to study the effectiveness of our score calibration, we can compare the MMB and CMB.
As can be seen, score calibration improves both indoor and outdoor scene predictions for i-vector systems.
\subsection{VGG-net vs I-Vectors}
\label{subsec:cnn_iv}
Looking at Table~\ref{tab:class_wise}, we can compare the performance of our best i-vector system with our VGG-net.
Comparing CMB column with VGG column shows that for some of the classes, i-vector system performs better 
than VGG-net and for some other VGG-net achieves better prediction results.
By looking at the HYB column we can observe that for most of the classes, the hybrid method performs better than both the i-vector and VGG-net.
Only in the \emph{City} and \emph{Tram} classes the performance of the hybrid system is not better than VGG-net CNN and i-vector.
Although it is not worse than the average of them in those cases, and its overall performance is better than both.
\subsection{Final thoughts on our hybrid approach}
\label{subsec:fin_thought}
Our experimental results support our hypothesis that i-vectors and CNNs provide complementary information from acoustic scenes.
%By summarizing the discussions in Section~\ref{subsec:cnn_iv},
Hence, we can conclude that both of our feature learning method (CNN) and feature modeling based on engineered features (i-vector) system
are capable of capturing acoustic events, enabling them to achieve promising performances in ASC.
Studying the class-wise performances shows that each of these methods model the acoustic events differently from another.
We provided a solution for combining the two modeling approaches by first create probability-like scores from each method and further fuse the scores.
Our score fusion technique enables us to benefit from both methods, while minimizes the differences between the training and validation score distributions.
%This experimental result suggests that the information captured from different audio scenes by i-vector system and VGG-net differ from one another.
%Also since overall performance of the hybrid system is improved compared to the i-vector and VGG-net, we believe this shows that the i-vectors and CNNs are capturing complementary information from audio scenes.
%
%
%
% Section by Hamid
%
\section{Conclusion}
\label{sec:conclude}
In this paper, we investigated the parametrizations of MFCCs and 
provided a setup for MFCC extraction for ASC using i-vectors.
We further proposed a novel multi-channel i-vector extraction scheme for ASC which uses 
different channels of the audio and significantly improves the performance of i-vector systems for ASC.
%Finally, we used four different audio channels (left, right, average and difference) to extract i-vectors using tuned MFCCs.
We designed a VGG-style CNN architecture that achieves promising ASC results using spectrograms of audio segments.
%We further proposed a novel binaural i-vector extraction scheme using different channels of the audio.
%Finally after averaging scores of different channels, we used a score calibrations which improved the performance of our system. %by 
%integrating the power features learned by deep CNNs into our binaural i-vector system.
%We provided results supporting the performance gain of all of our proposed steps to improve the i-vector features for ASC.
We investigated the differences between the features modeled based on engineered features (i-vectors) and a feature learning method (CNN),
and showed they capture complementary information from acoustic scenes.
Finally, we proposed a hybrid ASC system by fusing our multi-channel i-vectors with our VGG-net.
Using our hybrid method, we achieved $1^{\text{st}}$ rank in the DCASE-2016 challenge~\cite{EghbalZadehdcase2016} 
and using our multi-channel i-vector system we ranked $2^{\text{nd}}$.
%\small
\bibliographystyle{IEEEbib}
\bibliography{refs}

\begin{thebibliography}{10}

\bibitem{valero2012gammatone}
X.~Valero and F.~Alias,
\newblock ``Gammatone cepst. coeffs: Biologically inspired features for
  non-speech audio classification,''
\newblock {\em Tran. on Mult.}, 2012.

\bibitem{dehak2011front}
N.~Dehak, P.~Kenny, R.~Dehak, P.~Dumouchel, and P.~Ouellet,
\newblock ``Front-end factor analysis for speaker verification,''
\newblock .

\bibitem{bisot2016acoustic}
V.~Bisot, R.~Serizel, S.~Essid, and G.~Richard,
\newblock ``Acoustic scene classification with matrix factorization for
  unsupervised feature learning,''
\newblock in {\em ICASSP}, 2016.

\bibitem{salamon2016deep}
J.~Salamon and J.~P. Bello,
\newblock ``Deep cnns and data augmentation for environmental sound
  classification,''
\newblock {\em arXiv preprint}, 2016.

\bibitem{zeinali2016deep}
H.~Zeinali, L.~Burget, H.~Sameti, O.~Glembek, and O.~Plchot,
\newblock ``Dnns and hidden markov models in i-vector-based text-dependent
  speaker verification,''
\newblock in {\em Odyssey Workshop}, 2016.

\bibitem{bahari2013accent}
M.~H. Bahari, R~Saeidi, and D.~Van~Leeuwen,
\newblock ``Accent recognition using i-vector, gaussian mean supervector and
  gaussian posterior probability supervector for spontaneous telephone
  speech,''
\newblock in {\em ICASSP}, 2013.

\bibitem{eghbalzISMIR2015ivecsim}
H.~Eghbal-zadeh, B.~Lehner, M.~Schedl, and G.~Widmer,
\newblock ``I-vectors for timbre-based music similarity and music artist
  classification,''
\newblock in {\em ISMIR}, 2015.

\bibitem{elizalde2013vector}
B.~Elizalde, H.~Lei, G.~Friedland, and N.~Peters,
\newblock ``An i-vector based approach for audio scene detection,''
\newblock {\em DCASE workshop}, 2013.

\bibitem{kenny2005joint}
P.~Kenny,
\newblock ``Joint factor analysis of speaker and session variability: Theory
  and algorithms,''
\newblock Tech. {R}ep., 2005.

\bibitem{garcia2011analysis}
D.~Garcia-Romero and C.~Espy-Wilson,
\newblock ``Analysis of i-vector length normalization in speaker recognition
  systems.,''
\newblock in {\em INTERSPEECH}, 2011.

\bibitem{scholkopft1999fisher}
S.~Mika, G.~Ratsch, J.~Weston, B.~Sch{\"{o}}lkopf, and K.-R. Muller,
\newblock ``Fisher discriminant analysis with kernels,''
\newblock 1999.

\bibitem{hatch2006within}
A.~O. Hatch, S.~S. Kajarekar, and A.~Stolcke,
\newblock ``Within-class covariance normalization for svm-based speaker
  recognition.,''
\newblock in {\em INTERSPEECH}, 2006.

\bibitem{dehak2010cosine}
N.~Dehak, R.~Dehak, J.~R. Glass, D.~A. Reynolds, and P.~Kenny,
\newblock ``Cosine similarity scoring without score normalization
  techniques.,''
\newblock in {\em Odyssey workshop}, 2010.

\bibitem{lehner2011mfcc}
B.~Lehner, R.~Sonnleitner, and G.~Widmer,
\newblock ``{Towards Light-weight, Real-time-capable Singing Voice
  Detection},''
\newblock in {\em ISMIR}, 2013.

\bibitem{Voice99}
M.~Brookes,
\newblock ``{Voicebox: Speech Processing Toolbox for Matlab},'' Website, 1999.

\bibitem{lin2013network}
Min Lin, Qiang Chen, and Shuicheng Yan,
\newblock ``Network in network,''
\newblock {\em arXiv preprint}, 2013.

\bibitem{brummer2007focal}
N.~Br{\"u}mmer,
\newblock ``Focal multi-class: Toolkit for evaluation, fusion and calibration
  of multi-class recognition scores—tutorial and user manual—,''
\newblock Tech. {R}ep., 2007.

\bibitem{breiman1996bagging}
L.~Breiman,
\newblock ``Bagging predictors,''
\newblock {\em Machine learning}, 1996.

\bibitem{mesaros2016tut}
A.~Mesaros, T.~Heittola, and T.~Virtanen,
\newblock ``Tut database for acoustic scene classification and sound event
  detection,''
\newblock in {\em EUSIPCO}, 2016.

\bibitem{Bisot2016}
V.~Bisot, R.~Serizel, S.~Essid, and G.~Richard,
\newblock ``Supervised nonnegative matrix factorization for acoustic scene
  classification,''
\newblock Tech. {R}ep., DCASE2016 Challenge, 2016.

\bibitem{Valenti2016}
M.~Valenti, A.~Diment, G.~Parascandolo, S.~Squartini, and T.~Virtanen,
\newblock ``Acoustic scene classification using convolutional neural
  networks,''
\newblock Tech. {R}ep., DCASE2016 Challenge, 2016.

\bibitem{EghbalZadehdcase2016}
H.~Eghbal-Zadeh, B.~Lehner, M.~Dorfer, and G.~Widmer,
\newblock ``{CP-JKU} submissions for {DCASE}-2016: a hybrid approach using
  binaural i-vectors and deep cnns,''
\newblock Tech. {R}ep., DCASE2016 Challenge, 2016.

\end{thebibliography}
\end{sloppy}
\end{document}